\documentclass[aps,prd,twocolumn,superscriptaddress,showkeys]{revtex4-2}  

\usepackage{graphicx}
\usepackage{subfigure}
\usepackage{epstopdf}
\usepackage{rotating}
\usepackage{scalerel}
\usepackage[colorlinks,linkcolor=blue,anchorcolor=blue,citecolor=blue]{hyperref}

\usepackage{comment}

\usepackage{amsmath}
\usepackage{amsfonts}
\usepackage{amsthm}
\usepackage{amsbsy}
\usepackage{amssymb}
\usepackage{bm}
\usepackage{braket}
\usepackage{cancel}
\usepackage{mathtools}
\usepackage{units}
\usepackage{upgreek}
\usepackage{wasysym}
\usepackage[version-1-compatibility]{siunitx}
\newcommand{\electron}{e^-}
\newcommand{\positron}{e^+}
\newcommand{\epem}{\positron\electron}

\newcommand{\proton}{p}
\newcommand{\antiproton}{\bar{p}}
\newcommand{\ppbar}{\proton\antiproton}

\usepackage{dcolumn}
\usepackage{color}
\usepackage{overpic}
\usepackage{xspace}
\usepackage{textpos}
\usepackage[english]{babel}
\usepackage{accents}

\usepackage{lineno}

\newcommand{\jpsi}{J/\psi}
\newcommand{\pip}{\pi^+}
\newcommand{\pin}{\pi^-}
\newcommand{\gevcc}{\mathrm{GeV}/c^2}

\newcommand{\gev}{\mathrm{GeV}}
\newcommand{\mev}{\mathrm{MeV}}
\newcommand{\pio}{\pi^0}
\newcommand{\chisq}{\chi^2}
\newcommand{\X}{\chi_{c1}(3872)}
\newcommand{\Yl}{\psi(4220)}
\newcommand{\Yh}{\psi(4390)}
\newcommand{\Zc}{Z_c(3900)}
\newcommand{\isr}{\gamma_{\strut\mathrm{ISR}}}
\newcommand{\pipi}{\pi^+\pi^-}
\newcommand{\ee}{e^+e^-}

\newcommand{\etal}{{\it et al. }}

\begin{document}



\title{Cross section measurement of $\boldmath{e^+e^- \to p\bar{p}\eta}$ and $\boldmath{e^+e^- \to p\bar{p}\omega}$ at center-of-mass energies between 3.773 GeV and 4.6 GeV}

\author{

M.~Ablikim$^{1}$, M.~N.~Achasov$^{10,c}$, P.~Adlarson$^{64}$, S. ~Ahmed$^{15}$, M.~Albrecht$^{4}$, A.~Amoroso$^{63A,63C}$, Q.~An$^{60,48}$, ~Anita$^{21}$, Y.~Bai$^{47}$, O.~Bakina$^{29}$
% ...
\\
\vspace{0.2cm}
(BESIII Collaboration)\\
\vspace{0.2cm}
{\it
  $^{1}$ Institute of High Energy Physics, Beijing 100049, People's Republic of China\\
% ...
$^{4}$ Bochum Ruhr-University, D-44780 Bochum, Germany\\
% ...
$^{10}$ G.I. Budker Institute of Nuclear Physics SB RAS (BINP), Novosibirsk 630090, Russia\\
% ...
$^{15}$ Helmholtz Institute Mainz, Johann-Joachim-Becher-Weg 45, D-55099 Mainz, Germany\\
% ...
$^{21}$ Indian Institute of Technology Madras, Chennai 600036, India\\
% ...
$^{29}$ Joint Institute for Nuclear Research, 141980 Dubna, Moscow region, Russia\\
% ...
$^{47}$ Southeast University, Nanjing 211100, People's Republic of China\\
  $^{48}$ State Key Laboratory of Particle Detection and Electronics, Beijing 100049, Hefei 230026, People's Republic of China\\
% ...
$^{60}$ University of Science and Technology of China, Hefei 230026, People's Republic of China\\
% ...
$^{63}$ (A)University of Turin, I-10125, Turin, Italy; (B)University of Eastern Piedmont, I-15121, Alessandria, Italy; (C)INFN, I-10125, Turin, Italy\\
  $^{64}$ Uppsala University, Box 516, SE-75120 Uppsala, Sweden\\
% ...
$^{c}$ Also at the Novosibirsk State University, Novosibirsk, 630090, Russia\\
% ...
}
\vspace{0.4cm}
}

\date{\today}

\begin{abstract}
  Based on \SI{14.7}{fb^{-1}} of $\epem$ annihilation data collected with the BESIII detector at the BEPCII collider at 17 different center-of-mass energies between \SI{3.7730}{GeV} and \SI{4.5995}{GeV}, Born cross sections of the two processes $\epem \to \ppbar\eta$ and $\epem \to \ppbar\omega$ are measured for the first time.
No indication of resonant production through a vector state $V$ is observed, and 
upper limits on the Born cross sections of $\epem \to V \to \ppbar\eta$ and $\epem \to V \to \ppbar\omega$ at the \SI{90}{\percent} confidence level are calculated for a large parameter space in resonance masses and widths.
For the current world average parameters of the $\psi(4230)$ of $m=4.2187~\gevcc$ and $\Gamma=44~\mev$, we find upper limits on resonant production of the $\ppbar\eta$ and $\ppbar\omega$ final states of $7.5~\textrm{pb}$ and $10.4~\textrm{pb}$ at the $90\%$ CL, respectively.
\end{abstract}


\maketitle

\section{\label{sec:intro} Introduction}
In recent years, several unexpected states have been discovered in the charmonium sector. Notable examples are the $\X$ discovered by Belle \cite{Choi:2003ue}, the charged charmonium-like $\Zc$ discovered by BESIII \cite{Ablikim:2013mio} and the vector states $\Yl$ and $\Yh$, originally discovered by BaBar \cite{Aubert:2005rm} as a single broad resonance named $Y(4260)$ in the $\ee\to\isr \pipi \jpsi$ process, and later found to be two distinct states by BESIII~\cite{ref1}.
Charmonium-like vector states similar to the $\Yl$ and $\Yh$ were observed by BESIII in the processes $\epem \to \pipi h_c$~\cite{ref2}, $\pipi \psi(3686)$~\cite{ref3}, $D^0D^{*-}\pi^+$~\cite{ref4}, $\omega\chi_{c0}$~\cite{ref5} and $\eta\jpsi$~\cite{bes3jpsieta1, bes3jpsieta2, ref6}. However, so far no observations have been made of decays into light mesons or baryons for either the $\Yl$ or the $\Yh$.
Cross sections have been determined for the processes $\ee\to \ppbar \pi^0$~\cite{ref7}, $\phi\phi\phi$, $\phi\phi\omega$~\cite{ref8}, $p K_S^0 \bar{n} K^-$~\cite{ref9}, $K_S^0 K^\pm \pi^\mp$~\cite{ref10}, $K_S^0 K^\pm \pi^\mp \pi^0$ and $K_S^0 K^\pm \pi^\mp \eta$~\cite{ref11} without significant detection of resonant production.
Channels that include a proton anti-proton pair are especially interesting, since the partial width of decays of the type $V\to\ppbar h$, where $V$ is a vector-state in the charmonium region and $h$ is a light meson, can be related to the production cross section $\ppbar \to V\; h$~\cite{Lundborg:2005am}. In light of the upcoming PANDA experiment at the FAIR facility~\cite{ref12}, it is important to obtain the production cross sections of potentially exotic resonances in the charmonium sector. 

Multiple theoretical explanations have been offered concerning the nature of the $\Yl$ and $\Yh$ states. Possible interpretations include $D_1 \bar{D}$ molecules, hybrid charmonia or baryonium states, and their compatibility with experimental data has recently been discussed in detail in Ref.~\cite{brambilla}. Additional information is needed from experiments in order to discriminate between the different hypotheses. Thus, the search for new decay modes of the $\Yl$ and $\Yh$ is important.  

In this work, we report measurements of the Born cross sections of the processes $\ee\to\ppbar\eta$ and $\ee\to\ppbar\omega$ for data collected at 17 different center-of-mass energies between $\sqrt{s}=3.7730~\gev$ and 4.5995~$\gev$ with the BESIII detector. The center-of-mass energy dependence of both cross sections is investigated for possible resonant contributions $V\to\ppbar\eta$ and $V\to\ppbar\omega$.

\section{\label{sec:bes} BESIII Detector and Monte Carlo Simulations}

The BESIII detector is a magnetic spectrometer~\cite{Ablikim:2009aa} located at the Beijing Electron Positron Collider (BEPCII)~\cite{Yu:IPAC2016-TUYA01}.
The cylindrical core of the BESIII detector consists of a helium-based multilayer drift chamber (MDC), a plastic scintillator time-of-flight system (TOF), and a CsI(Tl) electromagnetic calorimeter (EMC), which are all enclosed in a superconducting solenoidal magnet providing a 1.0~T magnetic field.
The solenoid is supported by an octagonal flux-return yoke with resistive plate counter muon identifier modules interleaved with steel.
The acceptance of charged particles and photons is 93\% over the $4\pi$ solid angle.
The charged-particle momentum resolution at $1~{\rm GeV}/c$ is $0.5\%$, and the $dE/dx$ resolution is $6\%$ for the electrons from Bhabha scattering.
The EMC measures photon energies with a resolution of $2.5\%$ ($5\%$) at $1$~GeV in the barrel (end cap) region.
The time resolution of the TOF barrel part is 68~ps, while that of the end cap part is 110~ps.
The end cap TOF system was upgraded in 2015 with multi-gap resistive plate chamber technology, providing a time resolution of 60~ps~\cite{etof}.  This improvement affects data at 11 of the 17 center-of-mass energy points.  

A Monte Carlo (MC) simulation of the full BESIII detector, based on \textsc{geant}{\footnotesize 4}~\cite{geant4}, is used to optimize selection requirements, determine the product of detector acceptance and reconstruction efficiency and study and estimate possible background contributions. These simulations also account for the observed beam energy spread.

Dedicated simulations with $10^6$ events per center-of-mass energy of the signal processes $\ee\to\ppbar\eta$ and $\ee\to\ppbar\omega$ with subsequent decays $\eta\to\gamma\gamma$, $\eta\to\pip\pin\pi^0$,  $\omega\to\pip\pin\pi^0$ and $\pi^0\to\gamma\gamma$ are generated with the {\sc ConExc}~\cite{ref:conexc} generator, accounting for initial state radiation (ISR) and vacuum polarization (VP). The three different decay modes of the $\eta$ meson are weighted according to the respective branching fraction as given by the Particle Data Group (PDG) Ref.~\cite{pdg}.

In addition, an inclusive MC sample at a center-of-mass energy of $\sqrt{s}=4.1784~\gev$, corresponding to the dataset with the largest integrated luminosity (see Table~\ref{tab:xseceta}), is used to study potential background contributions. This sample includes open charm processes, ISR production of vector charmonium(-like) states and continuum $q\bar{q}$ (where $q$ is a $u,d,s$ quark) processes. Known decay modes are modeled with {\sc evtgen}~\cite{ref:evtgen} using branching fractions taken from the PDG~\cite{pdg}, whereas unknown processes are modeled by the {\sc lundcharm} model~\cite{ref:lundcharm}. Final state radiation (FSR) from charged final state particles is incorporated with the {\sc photos} package~\cite{photos}.
The inclusive MC sample at $\sqrt{s}=4.1784~\gev$ corresponds to 40 times the luminosity for the data at this center-of-mass energy. 

\section{\label{sec:sel}Event selection}

Two different final states are studied in this work, namely $\ppbar\gamma\gamma$ (for $\ee\to\ppbar \eta$ with $\eta\to\gamma\gamma$) and $\ppbar\pipi\gamma\gamma$ (for $\ee\to\ppbar\eta$ with $\eta\to\pipi\pi^0$ and $\ee\to\ppbar\omega$ with $\omega\to\pipi\pi^0$, both with a subsequent $\pi^0\to\gamma\gamma$ decay). In the analysis, the invariant mass distributions $m(\gamma\gamma)$ and $m(\pipi\pi^0)$ will be used to identify and quantify $\eta$ and $\omega$ contributions.
The polar angle $\theta$ of each charged track detected in the MDC has to satisfy $|\cos\theta|<0.93$, and its point of closest approach to the interaction point must be within $\pm 10$~cm in the beam direction and within $1$~cm in the plane perpendicular to the beam direction.
For particle identification (PID), the TOF information and the specific energy deposit $dE/dx$ in the MDC are combined to calculate a likelihood $P(h)$ for the particle hypotheses $h=\pi, K, p$. The particle type with the largest likelihood is assigned to each track. In addition, we require a minimum likelihood of $P(h)>10^{-5}$ to suppress background.

For photons, a minimum energy deposit in the calorimeter of $25~\mev$ in the barrel region ($|\cos\theta|<0.80$) or of $50~\mev$ in the end-cap regions ($0.86<|\cos\theta|<0.92$) is required. In addition, the time information from the shower in the calorimeter relative to the event start time has to be less than $700~\textrm{ns}$. Showers within $10^\circ$ of the impact point of any charged 
track are discarded.  
%

\begin{figure}[tbh!]
\begin{overpic}[width=0.4\textwidth]{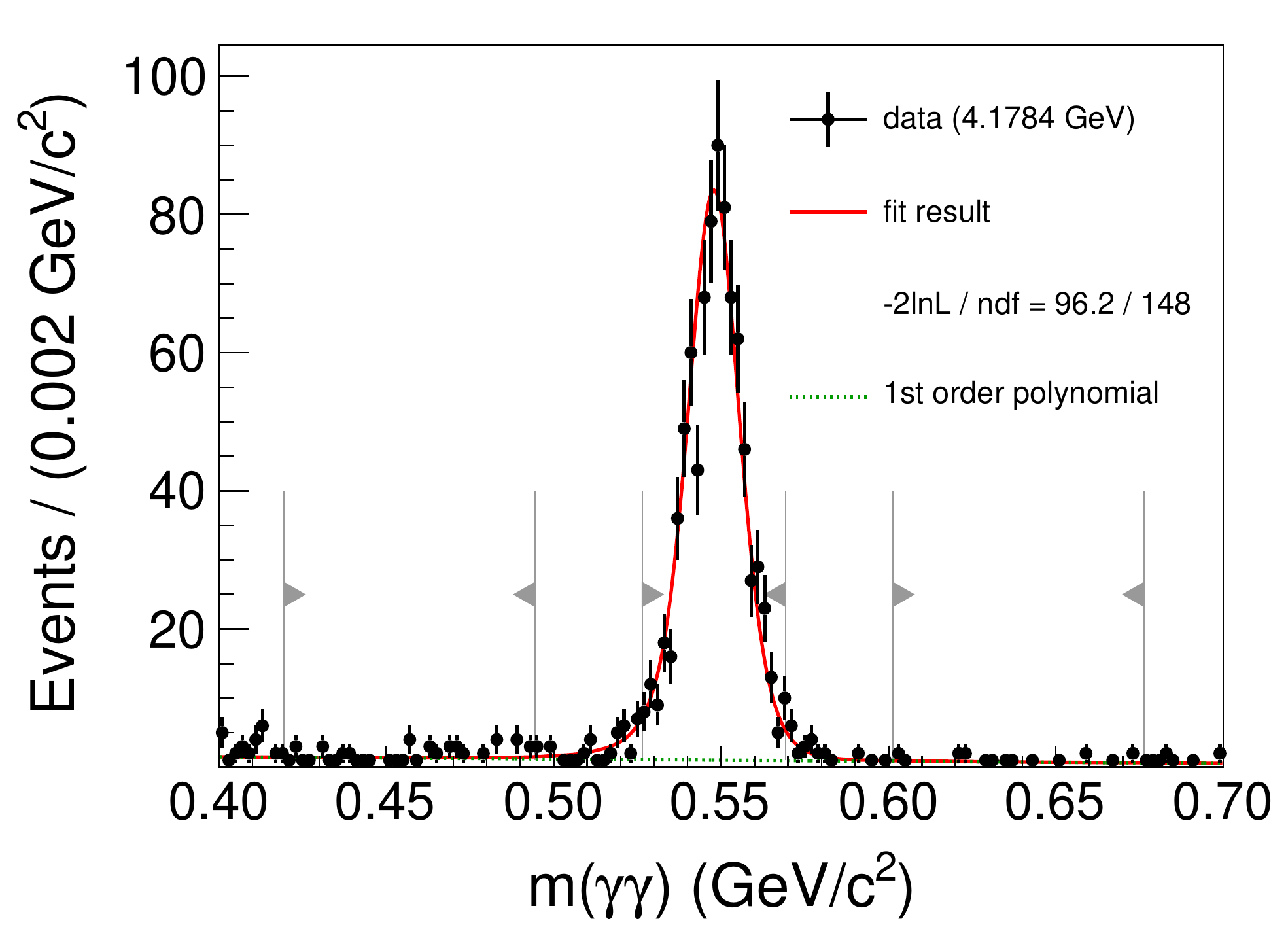}
\put(20,60){(a)}
\end{overpic}
\begin{overpic}[width=0.4\textwidth]{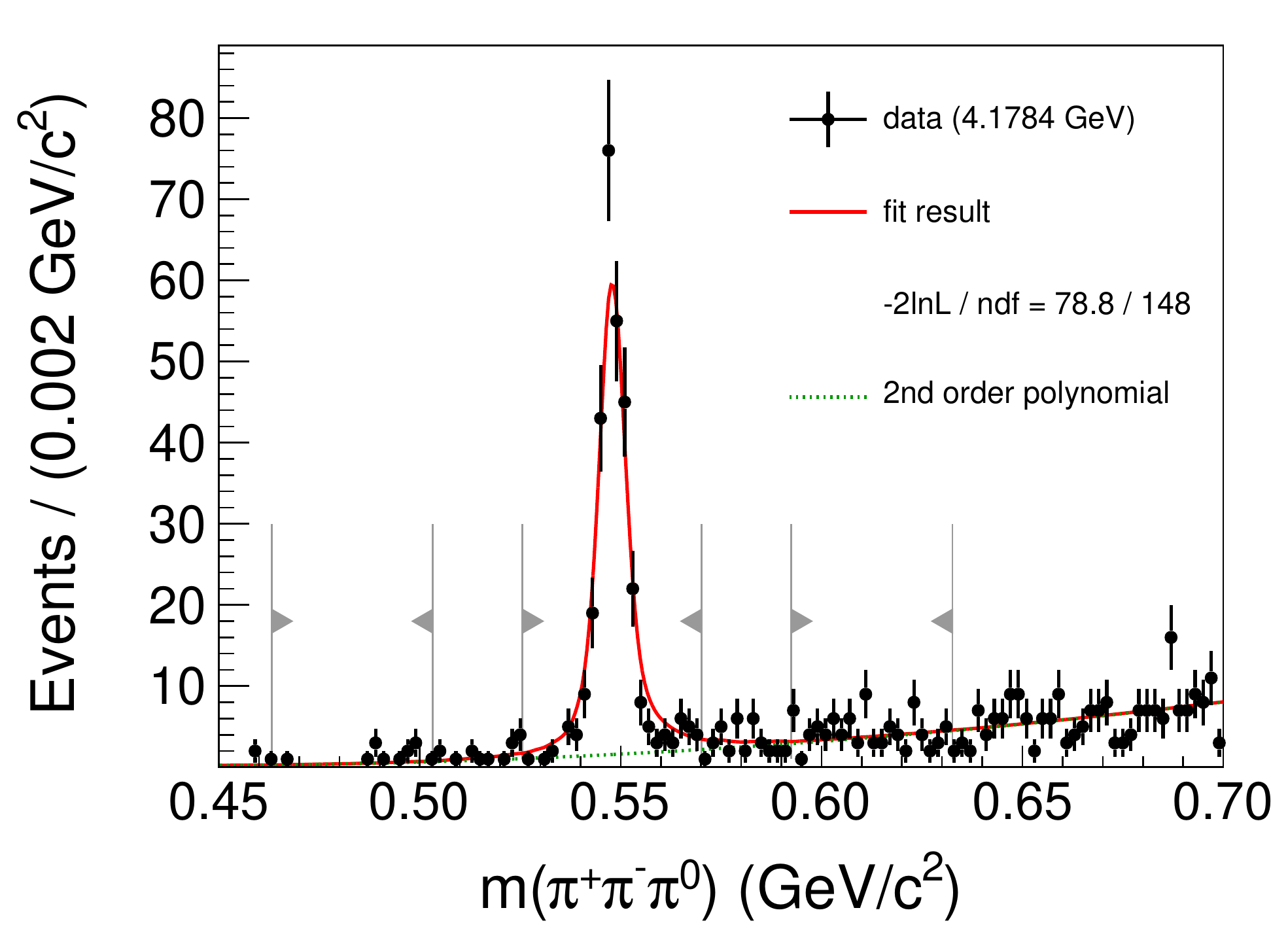}
\put(20,60){(b)}
\end{overpic}
\begin{overpic}[width=0.4\textwidth]{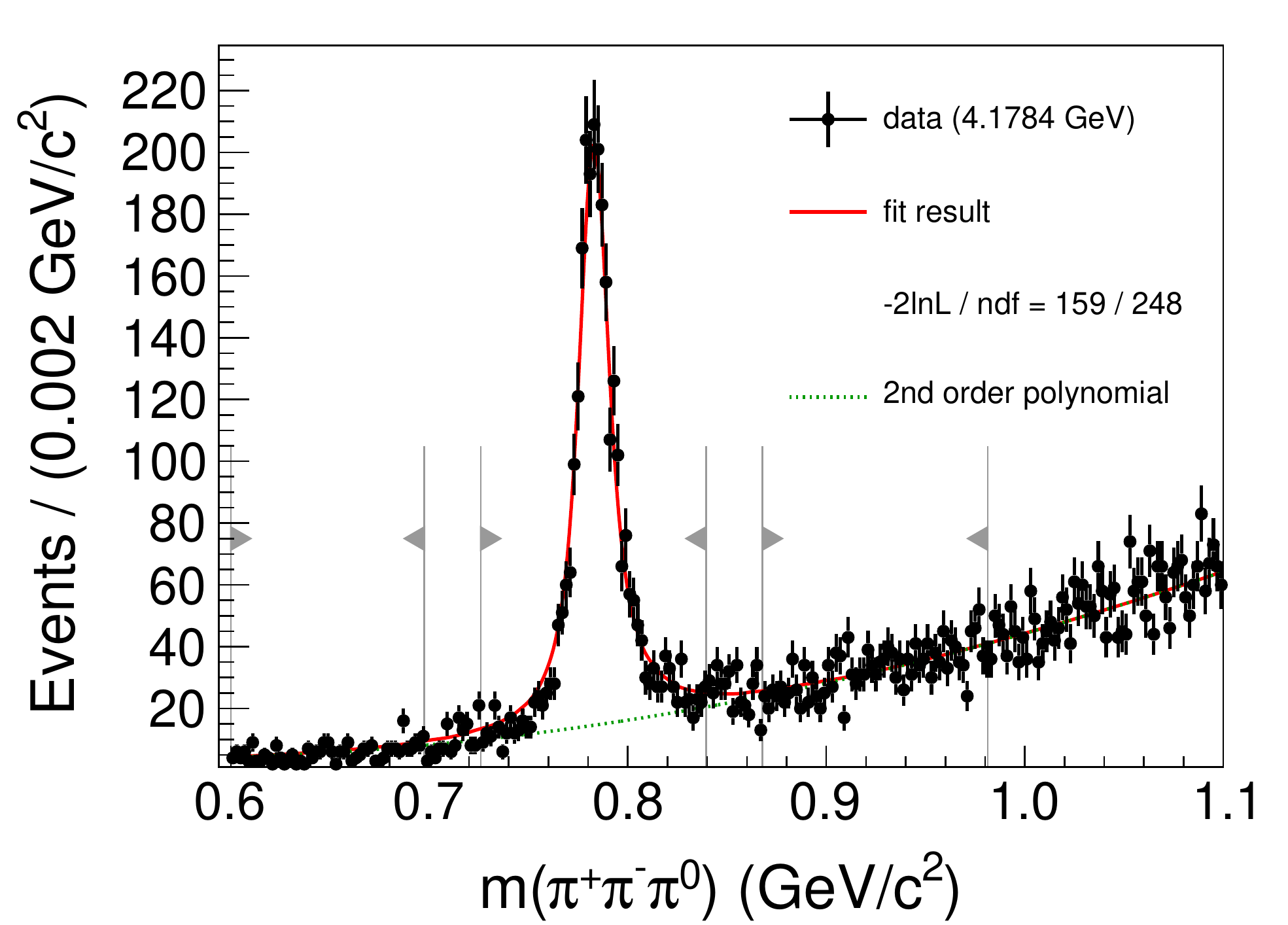}
\put(20,60){(c)}
\end{overpic}
 \caption{\label{fig:imfit} (Color online) Fits to the invariant mass of the (a) $\eta\to\gamma\gamma$, (b) $\eta\to\pipi\pi^0$ and (c) $\omega\to\pipi\pi^0$ systems. Black points represent data at the center-of-mass energy of $\sqrt{s}=4.1784~\gev$, full (red) lines represent the total fit result and short-dashed (green) lines show the background contribution. The gray markers indicate signal and sideband regions.}
\end{figure}

Only events containing exactly one good proton and one good anti-proton candidate (and exactly one good $\pi^+$ and one good $\pi^-$ candidate in case of the $\ppbar\pipi\gamma\gamma$ final state) and at least two good photon candidates are retained. A four- (five-)constraint kinematic fit is performed to the $\ppbar\gamma\gamma$ ($\ppbar\pipi\pi^0$ with $\pi^0\to\gamma\gamma$) hypothesis requiring four-momentum conservation between initial and final states and, if relevant, an additional mass constraint for the $\pi^0\to\gamma\gamma$ decay. If more than one $\gamma \gamma$ combination in an event satisfies the above requirements, only the combination with the lowest kinematic fit $\chi^2$ is kept for further analysis. 
The resulting invariant mass spectra for the decays $\eta\to\gamma\gamma$, $\eta\to\pipi\pi^0$ and $\omega\to\pipi\pi^0$ are displayed for the data at $\sqrt{s}=4.1784~\gev$ in Fig.~\ref{fig:imfit}.
According to the inclusive MC sample, backgrounds containing tracks misidentified as (anti-)proton candidates are very rare and can be neglected. In the case of the $\ppbar\gamma\gamma$ final state, the main background channels are the processes $\ee\to \ppbar$ 
, $ \ppbar\pi^0$, $\ppbar\omega$ with a subsequent $\omega\to\pi^0\gamma$ decay and $\ee\to\gamma_\textrm{ISR}\jpsi$ with a subsequent $\jpsi\to\ppbar$ decay. For the $\ppbar\pipi\gamma\gamma$ final state, the main background channels are $\ee\to\Delta\bar{\Delta}\pi$, $\ee\to p\bar{\Delta}\rho$, $\ee\to p\bar{\Delta}\pi\pi$ and $\ee\to \ppbar \rho \pi$ (the various possible charges of the $\Delta$, $\bar{\Delta}$, $\rho$ and $\pi$ and overall charge conjugation are taken into account), which can all lead to the signal final state. No peaking backgrounds in the invariant mass distributions $m(\gamma\gamma)$ and $m(\pipi\pi^0)$ are found.

The selection criteria with respect to the kinematic fit are optimized according to $\frac{s}{\sqrt{s+b}}$, where $s$ and $b$ are the number of signal and background events in the inclusive MC sample, that has been scaled to data, after requiring $\chi^2<\chi^2_\textrm{sel}$. In case of the $\ppbar\gamma\gamma$ final state, $\chi^2<30$ is found as an optimal selection condition, whereas for the other two reactions the background is dominated by other processes leading to the $\ppbar\pipi\pi^0$ system so that no requirement is made on the $\chi^2$ of the kinematic fit. This procedure is repeated with $s'$ and $b'$ being signal and background contributions determined directly from data using the $\eta$ ($\omega$) signal- and sidebands. The resulting selection condition agrees with the one determined from the inclusive MC sample within statistical uncertainties.
As both the data and the inclusive MC samples at the other center-of-mass energies are significantly smaller than those at $\sqrt{s}=4.1784~\gev$,
we do not repeat this optimization, and the requirement $\chi^2<30$ for the $\ppbar\gamma\gamma$ final state is applied for all center-of-mass energies.

The number of signal events is determined from a fit to the invariant mass spectra (see Fig.~\ref{fig:imfit}).
In the fit, the signal is described by a shape determined from signal MC simulations convolved with a Gaussian function in order to account for a possible underestimation of the mass resolution in MC simulation. The background is described by first- 
and second-order 
polynomial functions in the $\eta\to\gamma\gamma$ and $\eta(\omega)\to\pipi\pi^0$ channels, respectively. 
In the first step, we perform a global binned maximum likelihood fit to the sum of the data at all center-of-mass energies, determining a width of the Gaussian smearing functions as well as a background shape for each of the three channels. 
A binned maximum likelihood fit is then performed to each dataset and channel individually, with only the signal and background yields as free parameters.  
The signal and background shape are fixed to the result of the global fit 
in order to obtain reliable results from all sample sizes.  
The number of signal events in channel $i$ at center-of-mass energy $\sqrt{s}$ is then defined as $N_{\textrm{sig}}^i(s) = n_\textrm{sr} - n_{\textrm{bg,sr}}$, where $n_\textrm{sr}$ is the number of events in the signal region (defined as the symmetric region around the nominal $\eta$ ($\omega$) meson mass containing $95\%$ of all signal events according to the signal shape) and $n_{\textrm{bg,sr}}$ is the number of background events in the signal region according to the polynomial background function (see Fig.~\ref{fig:imfit}).

\section{\label{sec:eff} Efficiency Determination}

We define the efficiency,  $\epsilon^i(s)$, according to
%
\begin{equation} \epsilon^i(s) = \frac{N^i_\textrm{acc}(s)}{N^i_{\textrm{gen}}(s)} \quad , \label{eq:acc} \end{equation}
\noindent where $N^i_{\textrm{acc}}(s)$ is the number of reconstructed signal events and $N^i_{\textrm{gen}}(s)$ is the total size of the signal MC sample in channel $i$ for a center-of-mass energy $\sqrt{s}$. If the efficiency is not constant over the full $n$-particle phase-space, Eq.~\ref{eq:acc} only holds if the signal MC sample properly reflects data in all relevant coordinates $\vec{x}=\{p_p,\theta_p,\phi_p,p_{\bar{p}},\theta_{\bar{p}},\phi_{\bar{p}}, ...\}$. 
Since the data distribution is a priori unknown, we will perform 
a partial wave analysis of the data in order to re-weight our MC sample.  

The isobar model \cite{isobar} is used in the partial wave analysis by decomposing the full $\ee\to\gamma^*\to\ppbar\eta$ and $\ee\to\gamma^*\to\ppbar\omega$ processes
into a sequence of two-body decays. Each two-body decay is described in the helicity formalism ~\cite{helicity}.
%
%
%
%
The $\eta$ meson is treated as a stable particle in the amplitude analysis, whereas the three-body decay of the $\omega$ meson is described by a three-body amplitude according to Ref.~\cite{maltepaper}.
Blatt-Weisskopf barrier factors~\cite{helicity} are used for both the production $\gamma^*\to ad$ and the two-body decay $a\to bc$ according to Ref.~\cite{maltepaper} .
The dynamical part of the amplitude is described by relativistic Breit-Wigner amplitudes.
%
%
Only the line-shape of the $\eta$ and $\omega$ mesons is not described in this way. Here we employ the signal MC simulations that are used for normalization in the partial wave analysis for the line-shapes.
For the two $\ee\to\ppbar\eta$ channels, we include as intermediate $p\bar{p}$ resonances the $\jpsi$ (using a Voigt distribution) and possible contributions of a $J^{PC}=1^{--}$ and a $J^{PC}=3^{--}$ resonance. 
We also include one $J^P = \frac{1}{2}^{-(+)}$ and one $J^P = \frac{3}{2}^{+(-)}$ resonance contribution for the $\eta p (\bar{p})$ system. 
For of the $\ee\to\ppbar\omega$ channel, only intermediate states that decay to the $\ppbar$ system are included. For each of the three possible quantum numbers $J^{PC}=0^{-+},~0^{++},~2^{++}$ we include a phase-space contribution that is constant as a function of $m(\ppbar)$ as well as two resonant contributions.
All masses and widths of intermediate states are free parameters in the fit, apart from the $\jpsi$ contribution, where mass and width are fixed to the values given in the PDG~\cite{pdg}.  
Note that the aim of this partial wave analysis is only to describe 
the data accurately enough to enable an accurate determination 
of the efficiency.  

The partial wave analysis is performed as an unbinned maximum likelihood fit using the software package {\sc PAWIAN}~\cite{pawian}. Details on likelihood construction in {\sc PAWIAN} can be found in Refs.~\cite{maltepaper, pawian, malte2}. The remaining background events underneath the $\eta$ ($\omega$) peaks are accounted for in the partial wave analysis by adding the sidebands defined in Fig.~\ref{fig:imfit} to the likelihood with negative weights in such a way that the combined sideband weight is equal to the integral of the background function in the signal region.

The data for the two decay channels $\eta\to\gamma\gamma$ and $\eta\to\pipi\pi^0$ are fitted simultaneously with all amplitudes fully constrained between the two channels apart from an overall scaling factor.
The results of the partial wave analysis for the different channels are displayed in Fig.~\ref{pwa1} for the high statistics data at a center-of-mass energy of $\sqrt{s}=4.1784~\gev$.
In general, the fits have a similar quality for all center-of-mass energies. For each energy point, we obtain event weights, $w(\vec{x})$, from the partial wave analysis as a function of the coordinates in the $n$-particle phase-space, and they are used to determine the efficiency $\epsilon^i(s)$ as
\begin{equation}
 \epsilon^i(s) = \frac{\sum\limits_{j=0}^{N^i_{\textrm{acc}}(s)} w(\vec{x}_j)}{\sum\limits_{j=0}^{N^i_\textrm{gen}(s)} w(\vec{x}_j)} \quad .
\end{equation}
The efficiencies obtained in this way are summarized in Tables~\ref{tab:xseceta} and~\ref{tab:xsecomega}.

\begin{widetext}
$\quad$
\begin{figure}[tbh!]
\begin{overpic}[width=0.48\textwidth]{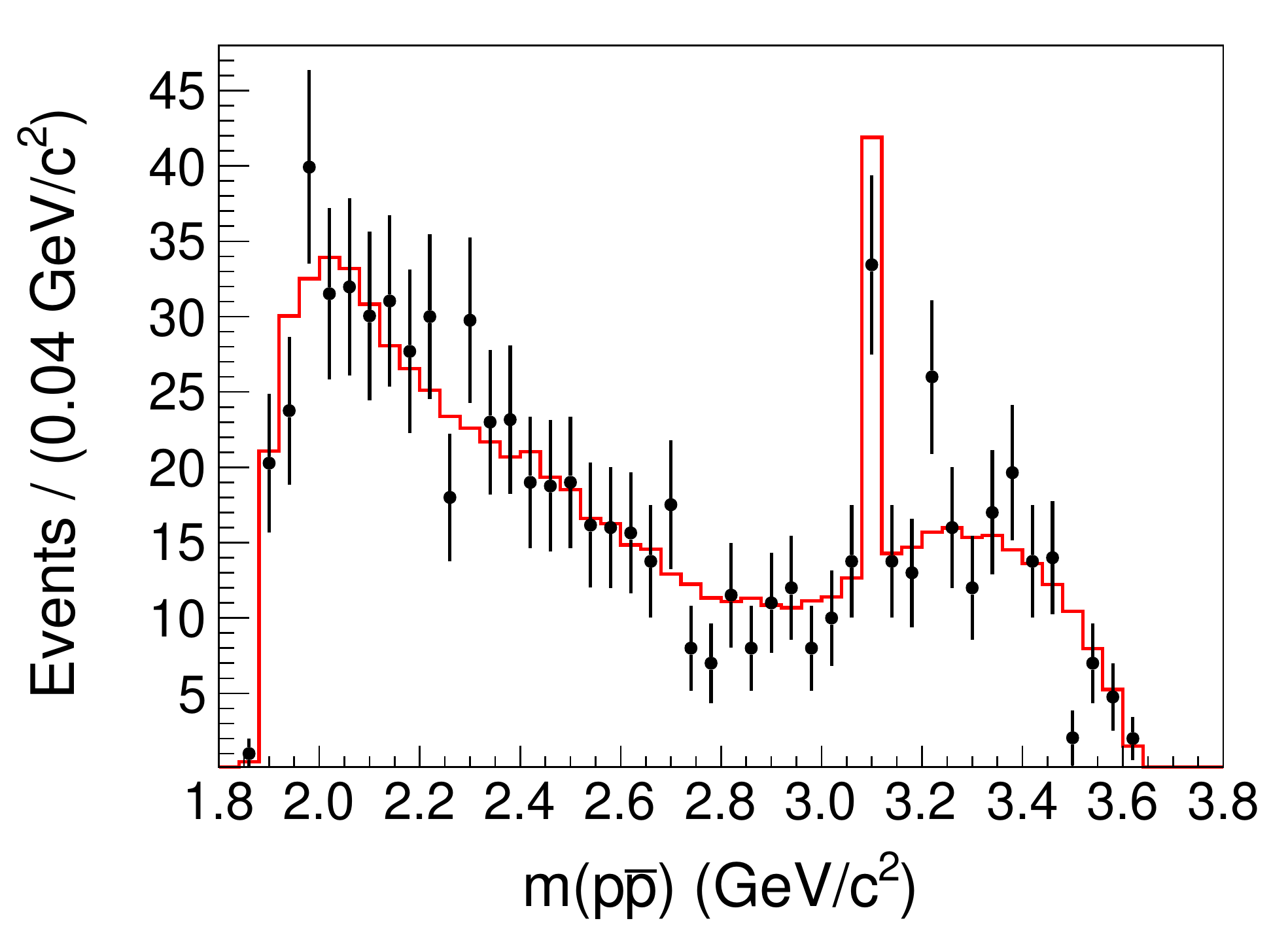}
\put(82,60){(a)}
\end{overpic}
\begin{overpic}[width=0.48\textwidth]{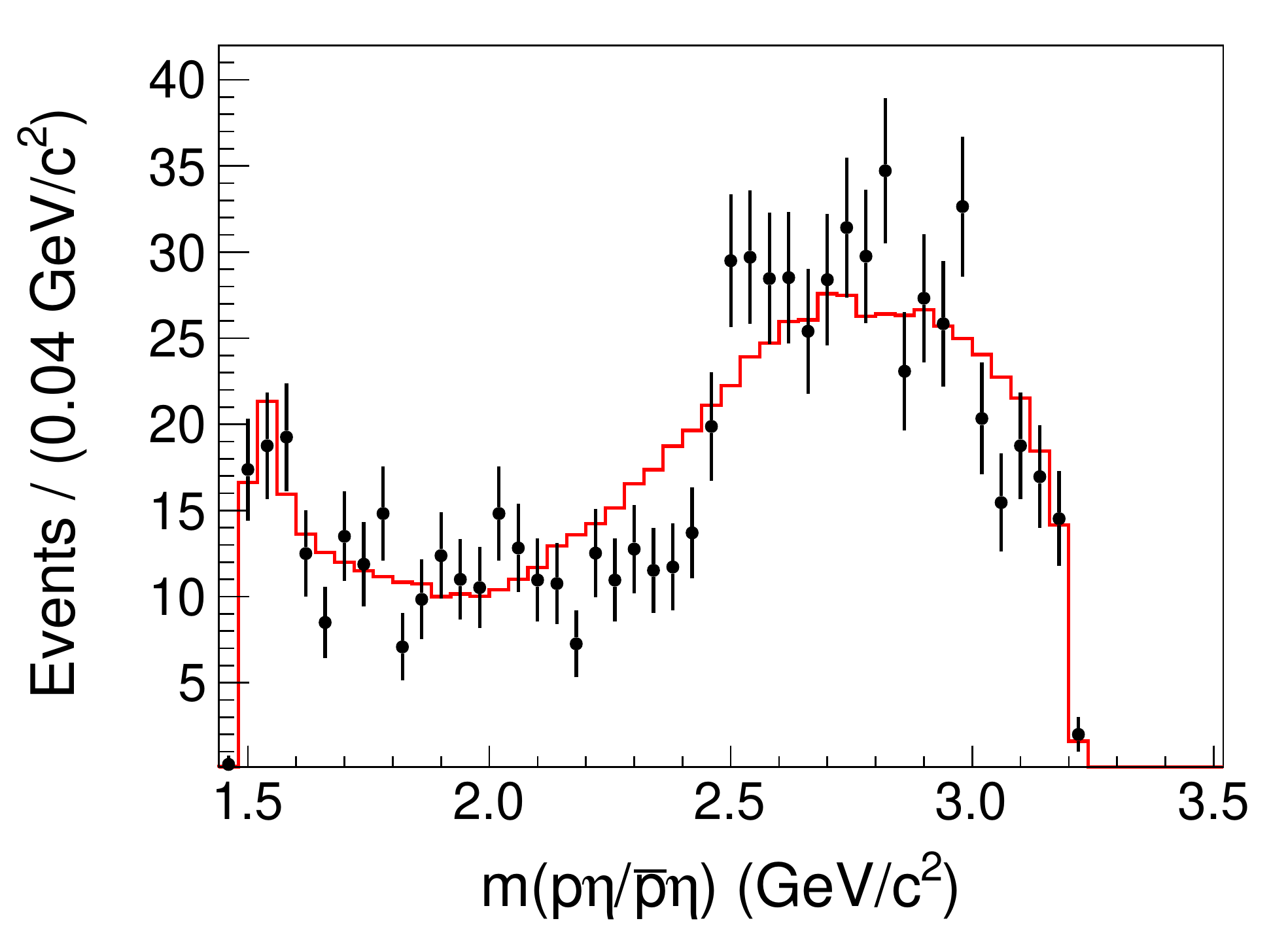}
\put(82,60){(b)}
\end{overpic}
\begin{overpic}[width=0.48\textwidth]{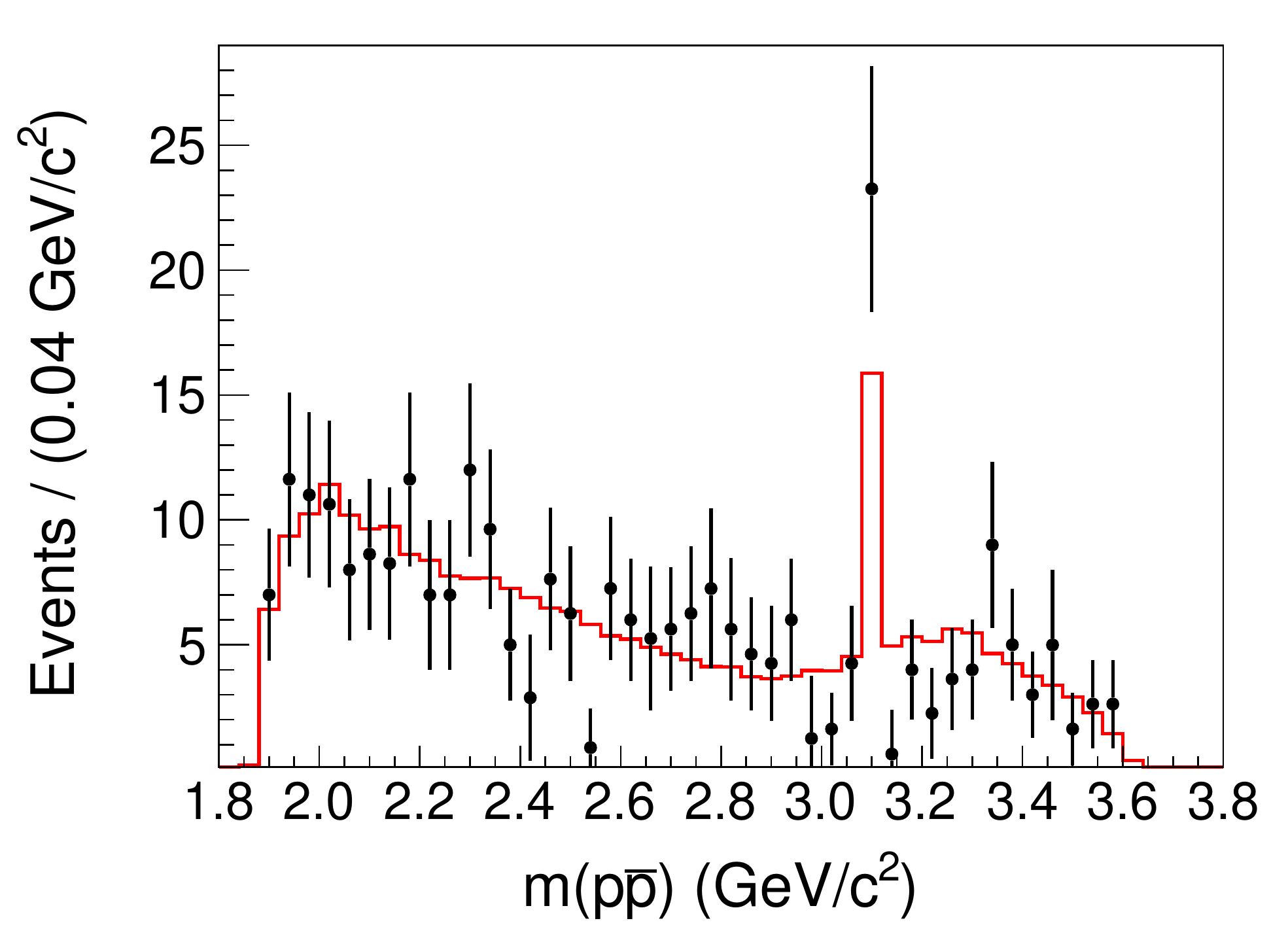}
\put(82,60){(c)}
\end{overpic}
\begin{overpic}[width=0.48\textwidth]{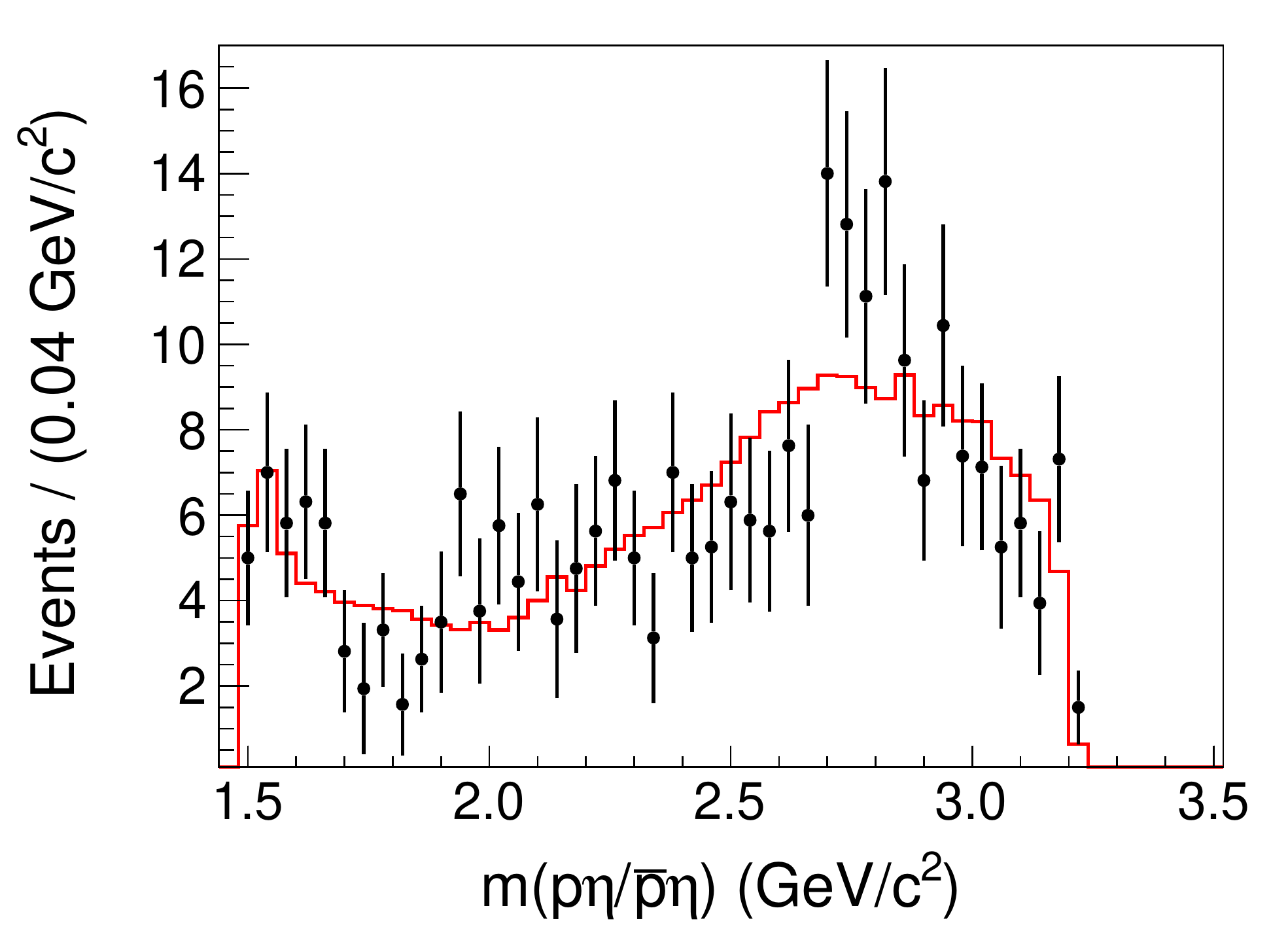}
\put(82,60){(d)}
\end{overpic}
\begin{overpic}[width=0.48\textwidth]{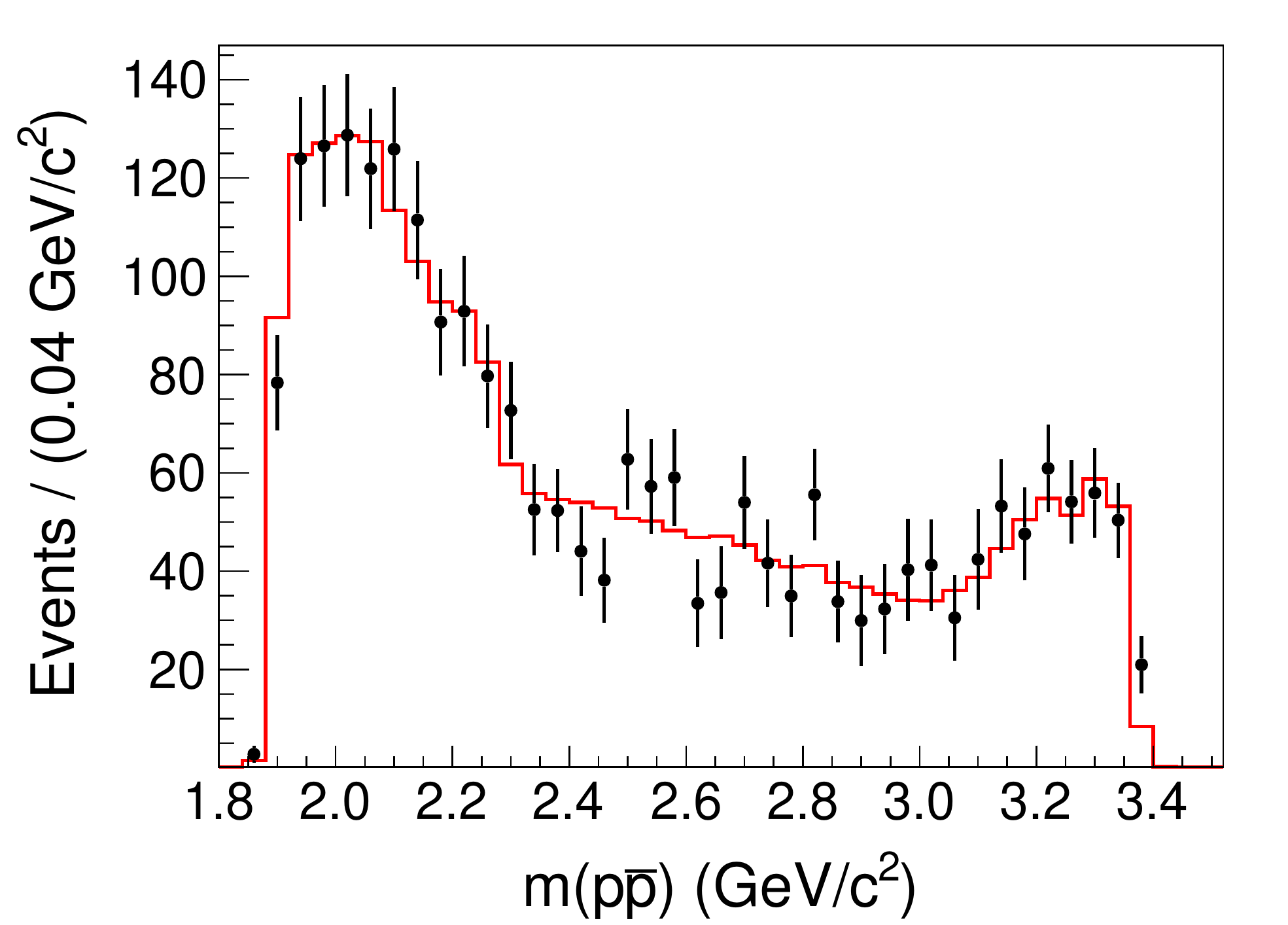}
\put(82,60){(e)}
\end{overpic}
\begin{overpic}[width=0.48\textwidth]{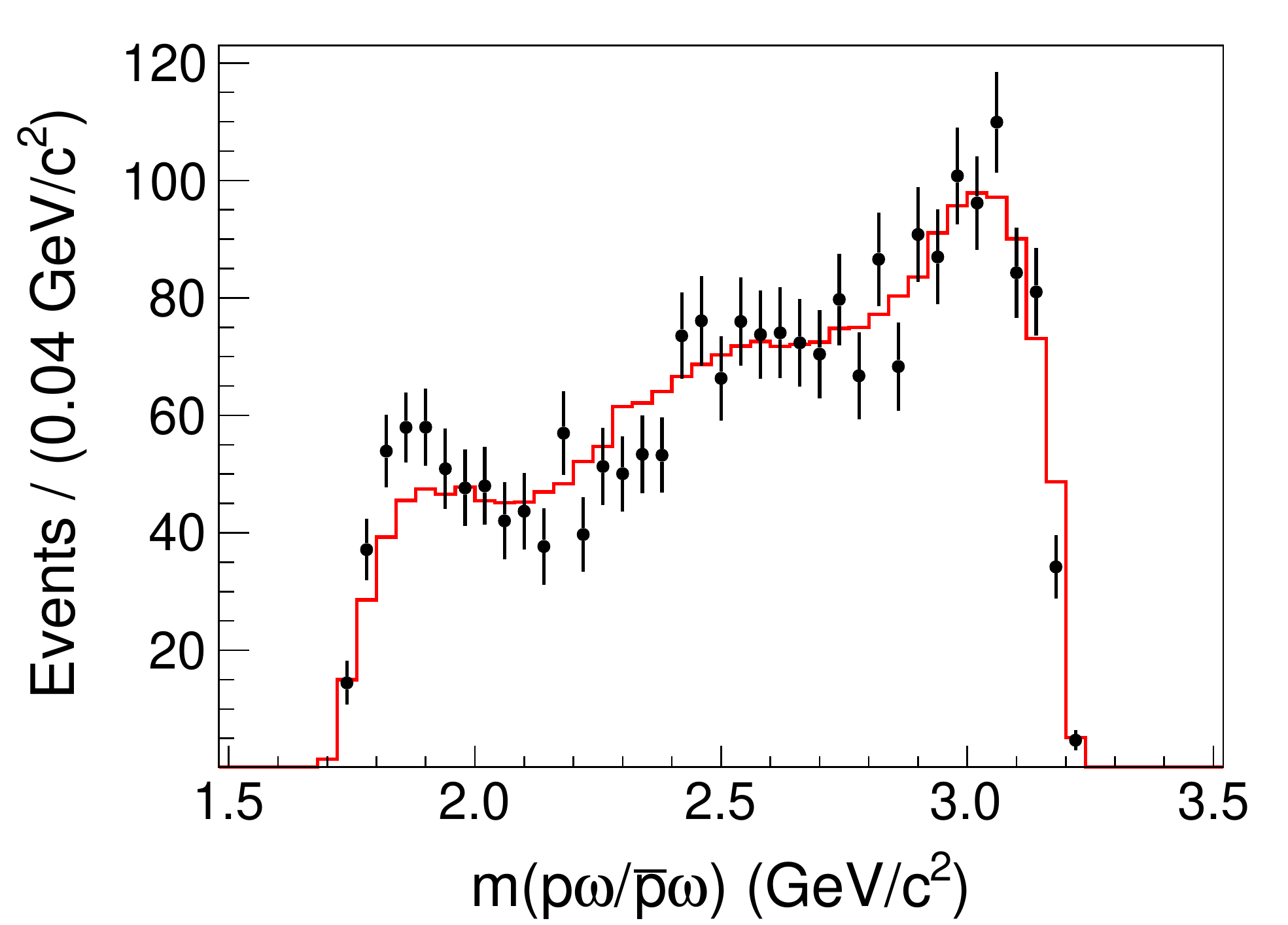}
\put(82,60){(f)}
\end{overpic}
 \caption{\label{pwa1} (Color online) Results of the partial wave analysis of the $\ee\to\ppbar\eta$ channel with subsequent $\eta\to\gamma\gamma$ (a - b) and $\eta\to\pi^+\pi^-\pi^0$ decays (c - d) and the $\ee\to\ppbar\omega$ process (e - f) for the data at a center-of-mass energy of $\sqrt{s}=4.1784~\gev$. The left column shows the invariant mass of the $\ppbar$ system, the right column the invariant mass of the $p\eta$ ($p\omega$) system. Black points correspond to data, full (red) lines show the result of the amplitude analysis.}
\end{figure}
\end{widetext}

\section{\label{sec:born} Determination of Born Cross Sections}

The Born cross section of the process $\ee\to\ppbar\eta$ ($\ee\to\ppbar\omega$) is given by
\begin{equation}
    \sigma_B(s) = \frac{N(s)}{L(s)\cdot (1+\delta_r(s))\cdot \frac{1}{|1-\Pi|^2} \cdot \epsilon(s)\cdot B} \quad , \label{eq:born}
\end{equation}
where $N(s)$ is the number of signal events observed in the data sample at center-of-mass energy $\sqrt{s}$, $L(s)$ is the corresponding integrated luminosity determined using Bhabha scattering~\cite{lumi}, $\delta_r(s)$ and $\frac{1}{|1-\Pi|^2}$ are corrections accounting for initial state radiation and vacuum polarization, $\epsilon(s)$ is the efficiency and $B$ is the product of branching ratios involved in the decay.
The correction $\frac{1}{|1-\Pi|^2}$ is calculated with the {\sc alphaQED} software package~\cite{alphaqed} with an accuracy of $0.5\%$. As initial state radiation depends on the shape of the cross section and can, in general, have an effect on the efficiency, it is treated in an iterative procedure starting from a flat energy dependence of the Born cross section $\sigma_B(s)$. We consider two successive iterations converged if $\kappa_i/\kappa_{i-1}=1$ within statistical uncertainties, where $\kappa(s)=\epsilon(s)\cdot (1+\delta_r(s))$ is the product of the efficiency and a radiative correction factor $1+\delta_r(s)$ obtained from the {\sc ConExc} MC generator.
The product of branching fractions is given by $B=Br(\eta\to\gamma\gamma)$ for the $\ee\to\ppbar\eta(\to\gamma\gamma)$ channel, $B=Br(\eta\to\pipi\pi^0)\,\cdot\,Br(\pi^0\to\gamma\gamma)$ for the $\ee\to\ppbar\eta(\to\pipi\pi^0)$ channel and $B=Br(\omega\to\pipi\pi^0)\,\cdot\, Br(\pi^0\to\gamma\gamma)$ for the $\ee\to\ppbar\omega(\to\pipi\pi^0)$ channel.
A combined Born cross section $\sigma_B$ is determined for the two different $\eta$ decay modes by using a weighted least squares method~\cite{agostini}.
%
%
%
%
The resulting cross sections are displayed in Fig.~\ref{fig:xsec} and all necessary values for their calculation are summarized in Tables ~\ref{tab:xseceta} and ~\ref{tab:xsecomega}.

\begin{widetext}
$\quad$
{\renewcommand{\arraystretch}{1.3}%
\begin{table}[h]
 \caption{\label{tab:xseceta} Summary of the Born cross sections $\sigma_B$ of the process $\ee\to\ppbar\eta$ for the datasets at different center-of-mass energies $\sqrt{s}$, integrated luminosity $L$, radiative corrections $1+\delta_r$, vacuum polarization correction $\frac{1}{|1-\Pi|^2}$, number of observed events $N_i$, efficiency $\epsilon_i$, and cross section $\sigma_i$ in the two channels (1) $\eta\to\gamma\gamma$ and (2) $\eta\to\pipi\pi^0$.}
 \begin{tabular}{c|c|c|c|l|c|c|l|c|c|c}
  \hline
  $\sqrt{s}$ (GeV) & $L$ (pb$^{-1}$) & $(1+\delta_r)$ & $\frac{1}{|1-\Pi|^2}$ & \multicolumn{1}{c|}{$N_1$} & $\varepsilon_1$ (\percent) & $\sigma_1$ (pb) & \multicolumn{1}{c|}{$N_2$} & $\varepsilon_2$ (\percent) & $\sigma_2$ (pb) & $\sigma_B$ (pb) \\
  \hline
  3.7730 & $   2931.8$ & 0.8993 & 1.057 & $      1521.9\ ^{+39.8}_{-38.2}$ & \SI{36.3+-0.1}{} & $3.82\ ^{+0.10}_{-0.10}$ & $   627.1\ ^{+   28.8}_{-   27.0}$ & \SI{24.8+-0.1}{} & $4.00\ ^{+0.18}_{-0.17}$ & $3.86\ ^{+0.09}_{-0.08}\ \pm 0.13$ \\
  3.8695 & $\ \,224.0$ & 0.9290 & 1.051 & $\ \,   106.8\ ^{+11.1}_{-\:\;9.5}$ & \SI{34.0+-0.1}{} & $3.65\ ^{+0.38}_{-0.32}$ & $\ \,39.9\ ^{+\:\;7.8}_{-\:\;6.1}$ & \SI{22.8+-0.1}{} & $3.53\ ^{+0.69}_{-0.54}$ & $3.61\ ^{+0.33}_{-0.28}\ \pm 0.13$ \\
  4.0076 & $\ \,482.0$ & 0.9553 & 1.044 & $\ \,   191.7\ ^{+14.7}_{-13.1}$ & \SI{33.1+-0.1}{} & $3.06\ ^{+0.23}_{-0.21}$ & $\ \,78.9\ ^{+   10.3}_{-\:\;8.6}$ & \SI{22.2+-0.1}{} & $3.27\ ^{+0.43}_{-0.36}$ & $3.11\ ^{+0.21}_{-0.18}\ \pm 0.11$ \\
  4.1784 & $   3189.0$ & 1.0398 & 1.054 & $\ \,   829.5\ ^{+29.6}_{-28.0}$ & \SI{30.0+-0.1}{} & $2.01\ ^{+0.07}_{-0.07}$ & $   282.8\ ^{+   18.3}_{-   16.7}$ & \SI{19.8+-0.1}{} & $1.81\ ^{+0.12}_{-0.11}$ & $1.95\ ^{+0.06}_{-0.06}\ \pm 0.07$ \\
  4.1888 & $\ \,524.6$ & 1.0034 & 1.056 & $\ \,   145.0\ ^{+13.0}_{-11.3}$ & \SI{30.8+-0.1}{} & $2.15\ ^{+0.19}_{-0.17}$ & $\ \,58.2\ ^{+\:\;9.3}_{-\:\;7.5}$ & \SI{20.3+-0.1}{} & $2.28\ ^{+0.36}_{-0.29}$ & $2.18\ ^{+0.17}_{-0.15}\ \pm 0.10$ \\
  4.1989 & $\ \,526.0$ & 1.0156 & 1.057 & $\ \,   133.9\ ^{+12.6}_{-10.9}$ & \SI{30.7+-0.1}{} & $1.96\ ^{+0.18}_{-0.16}$ & $\ \,62.8\ ^{+\:\;8.8}_{-\:\;7.2}$ & \SI{20.4+-0.1}{} & $2.41\ ^{+0.34}_{-0.28}$ & $2.05\ ^{+0.16}_{-0.14}\ \pm 0.08$ \\
  4.2092 & $\ \,518.0$ & 1.0240 & 1.057 & $\ \,   136.2\ ^{+12.4}_{-10.8}$ & \SI{30.8+-0.1}{} & $2.00\ ^{+0.18}_{-0.16}$ & $\ \,47.3\ ^{+\:\;8.0}_{-\:\;6.4}$ & \SI{20.1+-0.1}{} & $1.86\ ^{+0.31}_{-0.25}$ & $1.96\ ^{+0.16}_{-0.13}\ \pm 0.08$ \\
  4.2187 & $\ \,514.6$ & 1.1701 & 1.056 & $\ \,   107.2\ ^{+11.5}_{-\:\;9.9}$ & \SI{28.1+-0.1}{} & $1.52\ ^{+0.16}_{-0.14}$ & $\ \,38.4\ ^{+\:\;7.4}_{-\:\;5.7}$ & \SI{18.8+-0.1}{} & $1.42\ ^{+0.27}_{-0.21}$ & $1.49\ ^{+0.14}_{-0.12}\ \pm 0.07$ \\
  4.2263 & $   1056.4$ & 1.0228 & 1.056 & $\ \,   255.1\ ^{+16.7}_{-15.1}$ & \SI{30.7+-0.1}{} & $1.85\ ^{+0.12}_{-0.11}$ & $\ \,96.9\ ^{+   11.1}_{-\:\;9.5}$ & \SI{20.0+-0.1}{} & $1.88\ ^{+0.22}_{-0.18}$ & $1.85\ ^{+0.11}_{-0.09}\ \pm 0.07$ \\
  4.2357 & $\ \,530.3$ & 1.0569 & 1.055 & $\ \,   121.2\ ^{+12.0}_{-10.4}$ & \SI{29.9+-0.1}{} & $1.74\ ^{+0.17}_{-0.15}$ & $\ \,47.3\ ^{+\:\;7.9}_{-\:\;6.3}$ & \SI{19.8+-0.1}{} & $1.78\ ^{+0.30}_{-0.24}$ & $1.75\ ^{+0.15}_{-0.13}\ \pm 0.06$ \\
  4.2438 & $\ \,538.1$ & 1.0464 & 1.056 & $\ \,   120.2\ ^{+11.7}_{-10.1}$ & \SI{29.5+-0.1}{} & $1.74\ ^{+0.17}_{-0.15}$ & $\ \,46.7\ ^{+\:\;8.0}_{-\:\;6.4}$ & \SI{19.3+-0.1}{} & $1.80\ ^{+0.31}_{-0.25}$ & $1.75\ ^{+0.15}_{-0.13}\ \pm 0.07$ \\
  4.2580 & $\ \,828.4$ & 1.0536 & 1.053 & $\ \,   189.0\ ^{+14.6}_{-13.0}$ & \SI{30.7+-0.1}{} & $1.70\ ^{+0.13}_{-0.12}$ & $\ \,61.1\ ^{+\:\;9.0}_{-\:\;7.4}$ & \SI{20.0+-0.1}{} & $1.47\ ^{+0.22}_{-0.18}$ & $1.63\ ^{+0.11}_{-0.10}\ \pm 0.07$ \\
  4.2668 & $\ \,531.1$ & 1.0238 & 1.053 & $\ \,   132.4\ ^{+12.3}_{-10.7}$ & \SI{29.9+-0.1}{} & $1.96\ ^{+0.18}_{-0.16}$ & $\ \,42.1\ ^{+\:\;7.7}_{-\:\;6.1}$ & \SI{19.7+-0.1}{} & $1.65\ ^{+0.30}_{-0.24}$ & $1.87\ ^{+0.16}_{-0.13}\ \pm 0.07$ \\
  4.2777 & $\ \,175.7$ & 1.0463 & 1.053 & $\ \,\ \,39.1\ ^{+\ \,7.2}_{-\:\;5.6}$  & \SI{30.1+-0.1}{} & $1.71\ ^{+0.31}_{-0.24}$ & $\ \,15.0\ ^{+\:\;4.9}_{-\:\;3.3}$ & \SI{19.6+-0.1}{} & $1.75\ ^{+0.57}_{-0.38}$ & $1.71\ ^{+0.28}_{-0.21}\ \pm 0.09$ \\
  4.3583 & $\ \,543.9$ & 1.1749 & 1.051 & $\ \,\ \,80.7\ ^{+\ \,9.8}_{-\:\;8.2}$     & \SI{26.4+-0.1}{} & $1.15\ ^{+0.14}_{-0.12}$ & $\ \,26.7\ ^{+\:\;6.2}_{-\:\;4.6}$ & \SI{17.3+-0.1}{} & $1.02\ ^{+0.24}_{-0.18}$ & $1.12\ ^{+0.12}_{-0.10}\ \pm 0.04$ \\
  4.4156 & $   1043.9$ & 1.0714 & 1.052 & $\ \,   176.1\ ^{+14.3}_{-    12.6}$ & \SI{29.7+-0.1}{} & $1.28\ ^{+0.10}_{-0.09}$ & $\ \,57.7\ ^{+\:\;9.1}_{-\:\;7.4}$ & \SI{18.6+-0.1}{} & $1.16\ ^{+0.18}_{-0.15}$ & $1.25\ ^{+0.09}_{-0.08}\ \pm 0.05$ \\
  4.5995 & $\ \,586.9$ & 1.1439 & 1.055 & $\ \,\ \,80.1\ ^{+\ \,9.8}_{-\:\;8.2}$     & \SI{26.7+-0.1}{} & $1.07\ ^{+0.13}_{-0.11}$ & $\ \,15.5\ ^{+\:\;5.0}_{-\:\;3.4}$ & \SI{16.2+-0.1}{} & $0.60\ ^{+0.19}_{-0.13}$ & $0.92\ ^{+0.11}_{-0.08}\ \pm 0.03$ \\  \hline
  \end{tabular}
 \end{table}
 
\begin{table}[h]
 \caption{\label{tab:xsecomega} Summary of the Born cross sections $\sigma_B$ of the process $\ee\to\ppbar\omega$ for the datasets at different center-of-mass energies $\sqrt{s}$, integrated luminosity $L$, radiative corrections $1+\delta_r$, vacuum polarization correction $\frac{1}{|1-\Pi|^2}$, number of observed events $N$, and the efficiency $\epsilon$.}
 \begin{tabular}{c|c|c|c|c|c|c}
  \hline
  $\sqrt{s}$ (GeV) & $L$ ($pb^{-1})$ & $(1+\delta_r)$ & $\frac{1}{|1-\Pi|^2}$ & $N$ & $\varepsilon$ (\percent) & $\sigma_B$ (pb) \\
  \hline
  3.7730 & $   2931.8$ & 0.8978 & 1.057 & $   4623.7\ ^{+79.2}_{-77.4}$ & $30.9 \pm 0.1$ & $6.11\ ^{+0.10}_{-0.10}\ \pm 0.38$ \\
  3.8695 & $\ \,224.0$ & 0.9417 & 1.051 & $\ \,285.7\ ^{+19.9}_{-18.2}$ & $30.7 \pm 0.1$ & $4.76\ ^{+0.33}_{-0.30}\ \pm 0.30$ \\
  4.0076 & $\ \,482.0$ & 0.9832 & 1.044 & $\ \,485.4\ ^{+25.0}_{-23.3}$ & $30.8 \pm 0.1$ & $3.71\ ^{+0.19}_{-0.17}\ \pm 0.24$ \\
  4.1784 & $   3189.0$ & 1.0220 & 1.054 & $   2357.9\ ^{+53.9}_{-52.2}$ & $29.0 \pm 0.1$ & $2.68\ ^{+0.06}_{-0.06}\ \pm 0.17$ \\
  4.1888 & $\ \,524.6$ & 1.0406 & 1.056 & $\ \,370.4\ ^{+22.4}_{-20.7}$ & $28.2 \pm 0.1$ & $2.58\ ^{+0.16}_{-0.14}\ \pm 0.16$ \\
  4.1989 & $\ \,526.0$ & 1.0217 & 1.057 & $\ \,380.3\ ^{+22.1}_{-20.5}$ & $27.6 \pm 0.1$ & $2.76\ ^{+0.16}_{-0.15}\ \pm 0.17$ \\
  4.2092 & $\ \,518.0$ & 1.0393 & 1.057 & $\ \,355.7\ ^{+21.3}_{-19.6}$ & $28.3 \pm 0.1$ & $2.51\ ^{+0.15}_{-0.14}\ \pm 0.16$ \\
  4.2187 & $\ \,514.6$ & 1.0869 & 1.056 & $\ \,319.3\ ^{+20.4}_{-18.7}$ & $27.0 \pm 0.1$ & $2.27\ ^{+0.14}_{-0.13}\ \pm 0.15$ \\
  4.2263 & $   1056.4$ & 1.0145 & 1.056 & $\ \,733.8\ ^{+30.9}_{-29.2}$ & $28.6 \pm 0.1$ & $2.57\ ^{+0.11}_{-0.10}\ \pm 0.16$ \\
  4.2357 & $\ \,530.3$ & 1.0189 & 1.055 & $\ \,371.0\ ^{+22.0}_{-20.3}$ & $28.6 \pm 0.1$ & $2.57\ ^{+0.15}_{-0.14}\ \pm 0.17$ \\
  4.2438 & $\ \,538.1$ & 1.0463 & 1.056 & $\ \,353.4\ ^{+21.5}_{-19.8}$ & $28.7 \pm 0.1$ & $2.36\ ^{+0.14}_{-0.13}\ \pm 0.15$ \\
  4.2580 & $\ \,828.4$ & 1.0424 & 1.053 & $\ \,539.1\ ^{+26.4}_{-24.7}$ & $28.8 \pm 0.1$ & $2.34\ ^{+0.11}_{-0.11}\ \pm 0.15$ \\
  4.2668 & $\ \,531.1$ & 1.0287 & 1.053 & $\ \,347.0\ ^{+21.3}_{-19.6}$ & $26.3 \pm 0.1$ & $2.60\ ^{+0.16}_{-0.15}\ \pm 0.17$ \\
  4.2777 & $\ \,175.7$ & 1.0823 & 1.053 & $\ \,106.4\ ^{+12.8}_{-11.0}$ & $25.4 \pm 0.1$ & $2.37\ ^{+0.29}_{-0.25}\ \pm 0.15$ \\
  4.3583 & $\ \,543.9$ & 1.0569 & 1.051 & $\ \,300.3\ ^{+19.7}_{-18.0}$ & $26.2 \pm 0.1$ & $2.15\ ^{+0.14}_{-0.13}\ \pm 0.14$ \\
  4.4156 & $   1043.9$ & 1.0554 & 1.052 & $\ \,540.0\ ^{+26.0}_{-24.4}$ & $26.9 \pm 0.1$ & $1.96\ ^{+0.09}_{-0.09}\ \pm 0.13$ \\
  4.5995 & $\ \,586.9$ & 1.1230 & 1.055 & $\ \,210.4\ ^{+16.5}_{-14.8}$ & $25.1 \pm 0.1$ & $1.37\ ^{+0.11}_{-0.10}\ \pm 0.09$ \\
  \hline
  \end{tabular}
 \end{table}}
 
 \end{widetext}

\section{Systematic Uncertainties}

Various sources of systematic uncertainties contributing to the measurement of the $\ee\to\ppbar\eta$ and $\ee\to\ppbar\omega$ Born cross sections have been considered.

The uncertainty of the integrated luminosity determined using Bhabha scattering is $1\%$ \cite{lumi}. The systematic uncertainty of the tracking efficiency has been determined using a $\jpsi\to\ppbar\pipi$ control sample in Ref.~\cite{tracking} as $1\%$ per track. Similarly, systematic uncertainties of photon detection efficiencies have been studied using a $\jpsi\to\rho\pi$ control sample \cite{photon} and were found to be $1\%$ per photon.  For PID efficiency, a systematic uncertainty of $1\%$ per proton and $1\%$ per pion are taken from Ref.~\cite{ref7, pid}. 
For multiple particles, each of the track-finding, PID, and photon efficiency 
uncertainties are added linearly~\cite{tracking, photon, ref7, pid}. 
Uncertainties on the branching fractions are taken from the PDG~\cite{pdg}.
With regard to the kinematic fit, where a selection condition of $\chi^2<30$ is applied in case of the $\ee\to\ppbar\eta(\to\gamma\gamma)$ mode, the selection condition is varied between $\chi^2<5$ and $\chi^2<55$ in steps of $\delta\chi^2=5$ and the resulting Born cross section is determined and compared with the nominal value $R=\frac{\sigma_\textrm{step}}{\sigma_\textrm{nom}}$. We take the standard deviation of a weighted sample of the ratio $R$ as the systematic uncertainty due to potential differences in the $\chi^2$ distributions between data and MC simulation. Here, $1/\delta R$ is used as the weight, where $\delta R$ is the uncertainty taking into account the sizable correlation between the event samples.
The nominal symmetric signal region containing $95\%$ of the total signal is altered to a set of both smaller and larger signal regions and we determine the resulting Born cross sections.  As outlined above, we take the standard deviation of a sample of ratios $R$ weighted by the inverse of the statistical uncertainty as the systematic uncertainty resulting from the choice of signal region. 
For the background description, the polynomial shapes were increased by one order from the nominal first order polynomial used for the $\eta\to\gamma\gamma$ invariant mass spectrum, and the second order polynomial in the case of the $\eta\to\pipi\pi^0$ and $\omega\to\pipi\pi^0$ invariant mass spectra. The fits are then repeated and the difference to the nominal results is taken as a systematic uncertainty.
For the radiative correction factor, we performed five additional iterations and found no difference beyond the statistical uncertainty. This contribution to the systematic uncertainty is therefore neglected.

The systematic uncertainties are summarized in Table~\ref{tab:sys} for the data at a center-of-mass energy of $\sqrt{s}=4.1784~\gev$. The total systematic uncertainty is obtained by adding each contribution in quadrature. Correlated systematic uncertainties in the two $\ee\to \ppbar \eta$ channels are accounted for in the calculation of the combined Born cross section following Ref.~\cite{agostini}.

{\renewcommand{\arraystretch}{1.0}%
\begin{table}
 \caption{\label{tab:sys} Summary of systematic uncertainties in percent for the data at $\sqrt{s}=4.1784~\gev$.}
 \begin{tabular}{l|c|c|c}
  \hline
   & $\eta \to \gamma\gamma$ & $\eta \to \pip\pin\pio$ & $\omega \to \pip\pin\pio$ \\
  \hline
  Luminosity                & 1.0 & 1.0 & 1.0 \\
  Tracking efficiency       & 2.0 & 4.0 & 4.0 \\
  Photon detection          & 2.0 & 2.0 & 2.0 \\
  Particle Identification   & 2.0 & 4.0 & 4.0 \\
  Branching fraction        & 0.5 & 1.2 & 0.8 \\
  $\chisq$ cut              & 1.9 &     &     \\
  Signal region             & 0.6 & 0.8 & 0.8 \\
  Background description    & 0.7 & 1.5 & 0.3 \\
  \hline
  Total                     & 4.3 & 6.5 & 6.2 \\
  \hline
  \end{tabular}
 \end{table}

\section{Search For Resonant Contributions}

The final Born cross sections for the $\ee\to\ppbar\eta$ and $\ee\to\ppbar\omega$ processes are displayed in Fig.~\ref{fig:xsec}. In order to search for possible  $\ee\to V\to\ppbar\eta$ ($\ee\to V\to\ppbar\omega$) resonant contributions, we perform two different fits. In the first fit, only a non-resonant contribution of the type
\begin{equation}
    \sigma_{\textrm{nr}}(s) = \left(\frac{C}{\sqrt{s}}\right)^\lambda 
\end{equation}
defined in Ref.~\cite{ref7} is used. The second fit includes a single Breit-Wigner amplitude of the form
\begin{equation}
    A_\textrm{res}(s) =  
    A_V \, \left( \frac{m\Gamma}{s^2-m^2+im\Gamma} \right) 
\end{equation}
that is coherently added to the non-resonant term.

\begin{figure}[htb!]
\begin{overpic}[width=0.45\textwidth]{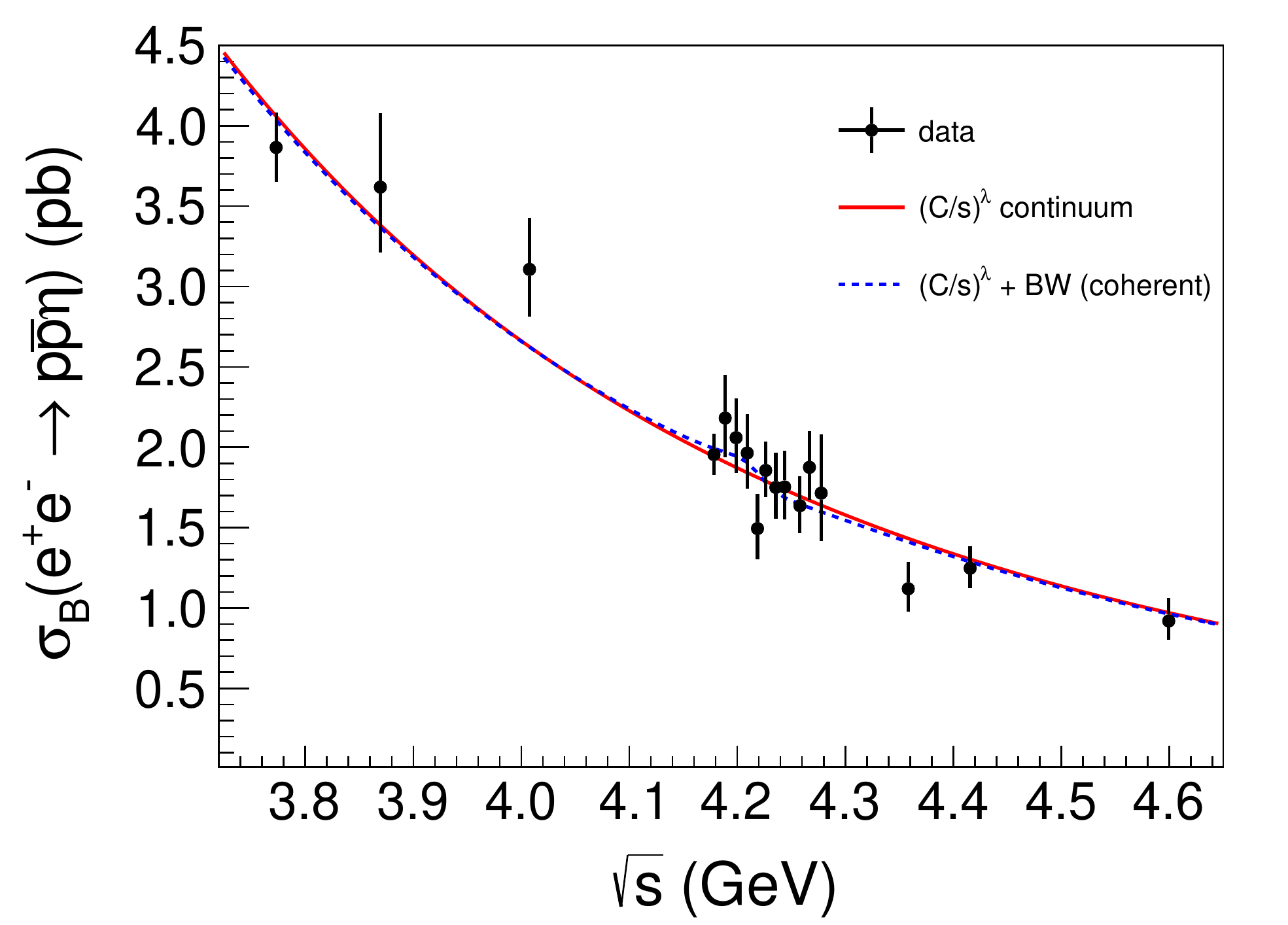}
\put(35,60){(a)}
\end{overpic}
\begin{overpic}[width=0.45\textwidth]{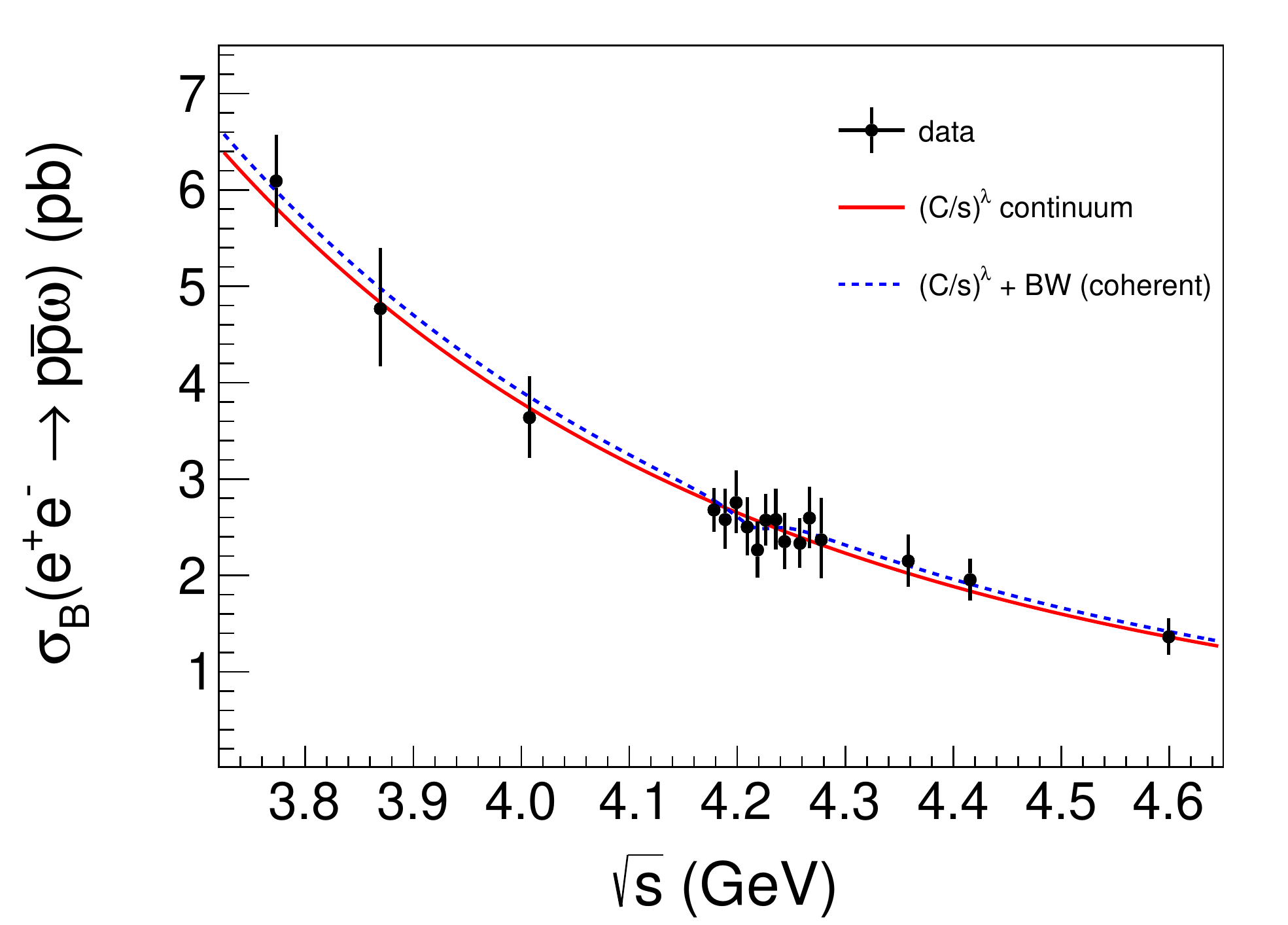}
\put(35,60){(b)}
\end{overpic}
\caption{\label{fig:xsec} (Color online) Born cross sections of the $\ee\to\ppbar\eta$ (a) and $\ee\to\ppbar\omega$ (b) processes as a function of the center-of-mass energy. Black points represent our result including both statistical and systematic uncertainties. The full (red) and long-dashed (blue) 
lines represent the fits using a continuum contribution and a Breit-Wigner coherently added to the continuum contribution, respectively. The fits displayed ($m=4.2187~\gevcc$ and $\Gamma=44~\mev$) are those for the current world average parameters of the $\psi(4230)$~\cite{pdg}.}
\end{figure}

Unbinned maximum likelihood fits are performed where the likelihood $L(x;\Theta)$ given the data $x$ and the fit parameters $\Theta$ is defined as the product $L(x;\Theta)=\prod_{i,j}L_{ij}(\Theta)$, where $L_{ij}$ is a set of likelihood functions, one each for each dataset $i$ and decay mode $j$. These likelihood functions are transformed such that they only depend on the expected number of signal events $N\equiv N_{ij}(\Theta)$ which can be calculated for each dataset according to Eq.~\ref{eq:born}. The likelihood $L_{ij}(N)$ is then obtained from data via a likelihood scan of the number of signal events in the invariant mass distributions of the meson decay systems. These likelihood scans are parameterized by asymmetric Gaussian distributions. Incorporating the systematic uncertainties of dataset $i$ and channel $j$, the likelihood is
\begin{eqnarray}
    L_{ij}(N) = \frac{1}{\sqrt{2\pi\left(\left(\frac{\sigma_L+\sigma_R}{2}\right)^2 + \sigma_\text{sys}^2\right)^2}} \cdot e^{-\frac{(N-\mu)^2}{2(\sigma_k^2+\sigma_\textrm{sys}^2)}} \nonumber \\
    \textrm{with}\quad \sigma_k = \left\{ \begin{matrix} \sigma_L~,~ N\leq \mu \\ \sigma_R~,~ N> \mu \end{matrix}\right.  \quad .
\end{eqnarray}
In the fit, all systematic uncertainties apart from the one on the branching ratio of the meson decays are considered uncorrelated between the different c.m. energies. While a correlation of a systematic uncertainty between two c.m. energies can not in general be ruled out, our assumption of a vanishing correlation leads to the most conservative upper limit estimation.
We find no evidence for a resonant contribution from the fits
and set upper limits at the $90\%$ confidence level. As the resonant contribution is added coherently, the fit finds two ambiguous solutions for constructive and destructive interference by construction. The upper limits are obtained by integrating $L(x;\Theta)=\prod_{i,j}L_{ij}(\Theta)$ according to
\begin{equation}
\frac{\int\limits_{-\infty}^{\sigma_V^{\textrm{UL}}} L(x,\Theta) \, \pi(\Theta) \,  d\sigma_V}{\int\limits_{-\infty}^{\infty} L(x,\Theta) \, \pi(\Theta) \, d\sigma_V} = 0.90 ~ ,
\end{equation}
where the prior $\pi(\Theta)$ is given by
\begin{equation}
\pi(\Theta) = \left\{ \begin{matrix} 1 ~ , ~ \sigma_V \geq 0 \\ 0 ~ , ~ \sigma_V < 0 \end{matrix} \right. \quad .
\end{equation}
The procedure outlined above is repeated  with a step size of $1~\mev$ for different masses $m$ in the range $4~\gevcc < m < 4.4~\gevcc$ and widths $\Gamma$ in the range $40~\mev < \Gamma < 300~\mev$ for a potential resonant contribution. The results are shown in Fig.~\ref{fig:ul}.
%

\begin{figure}[htb]
 \includegraphics[width=0.45\textwidth]{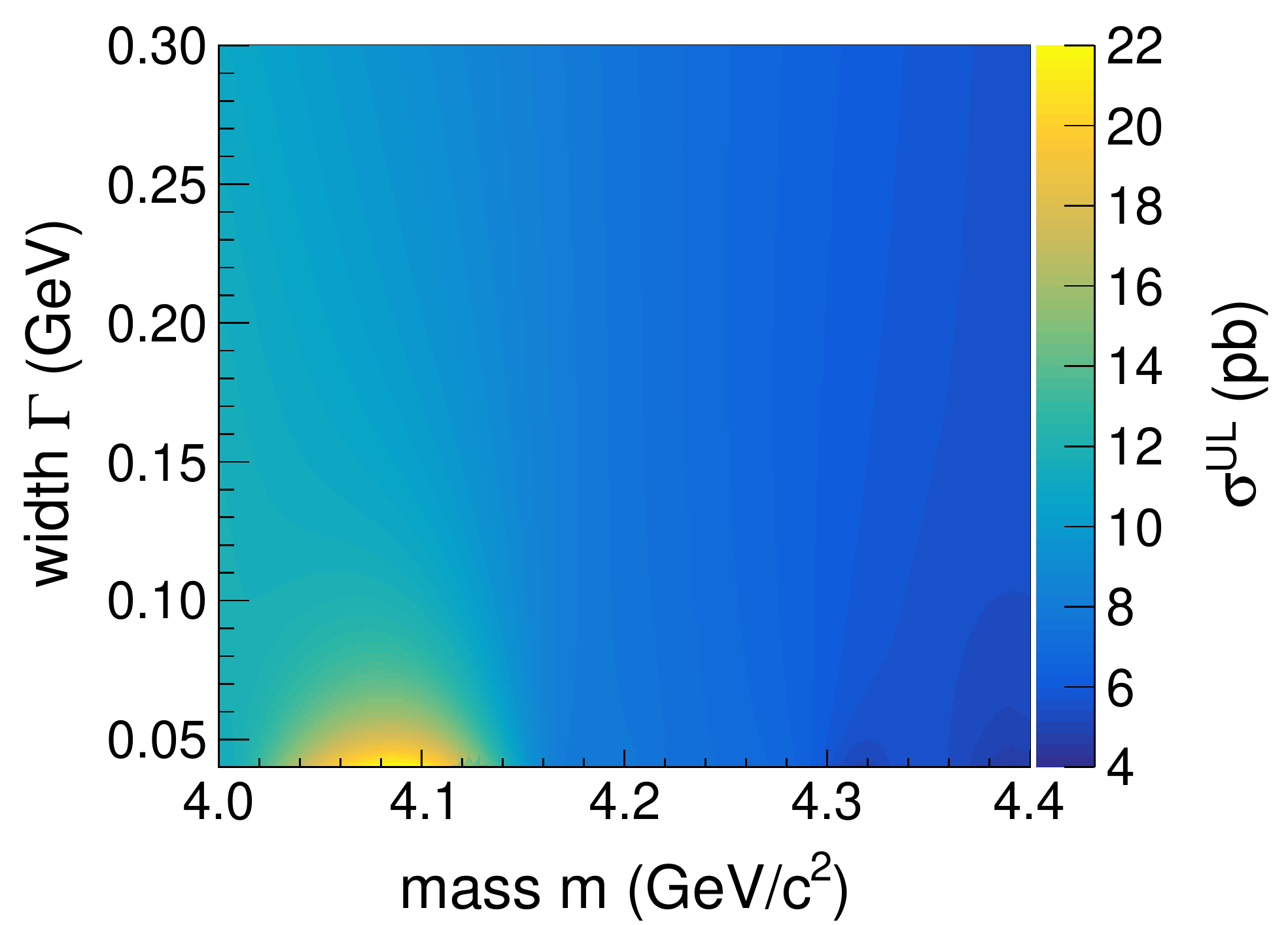}
 \includegraphics[width=0.45\textwidth]{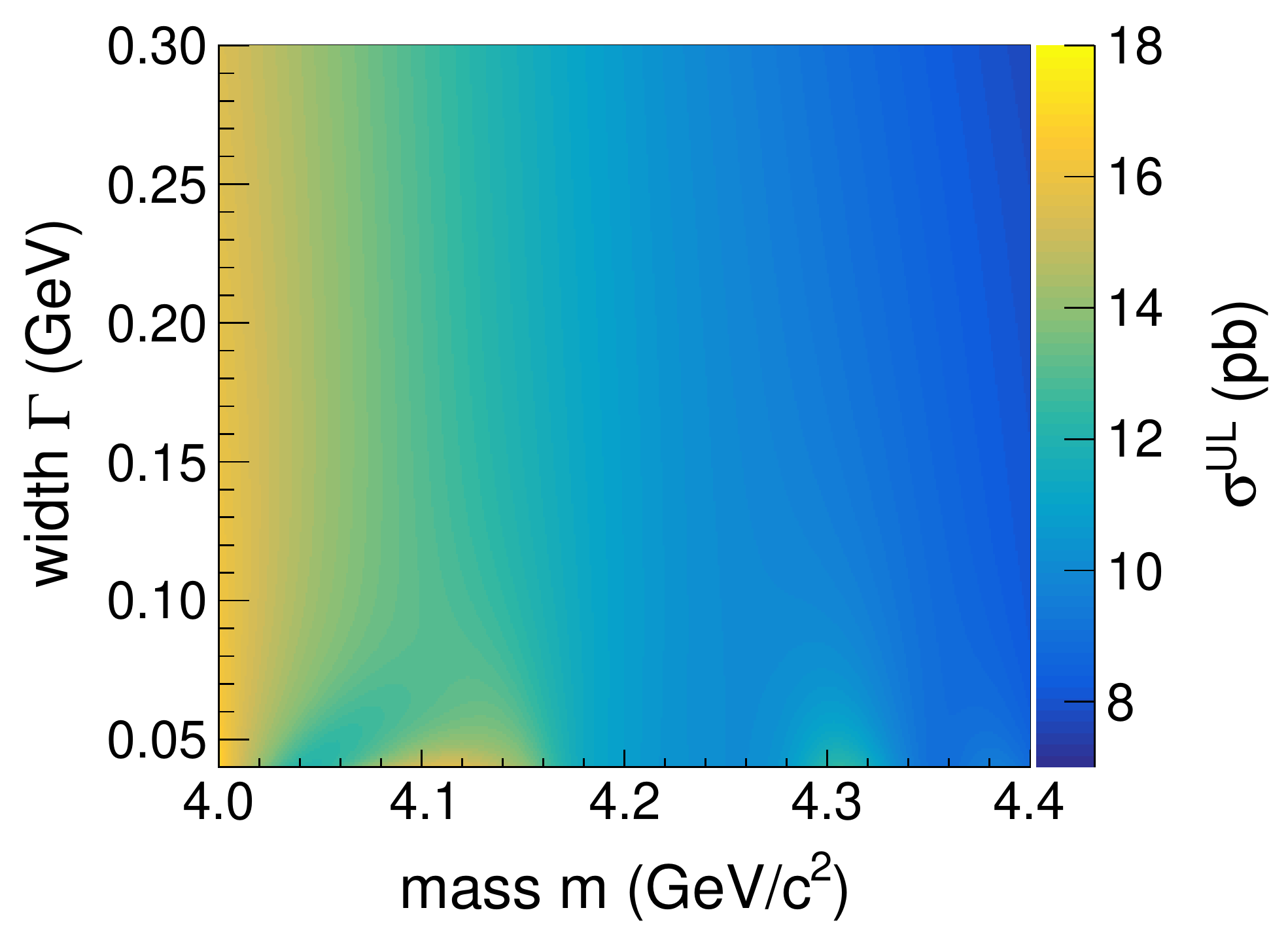}
 \caption{\label{fig:ul} (Color online) Upper limits on a possible resonant contribution with mass $m$ and width $\Gamma$
 for the two processes $\ee\to\ppbar\eta$ (top) and $\ee\to\ppbar\omega$ (bottom).}
\end{figure}
%
The most stringent upper limits are found for resonant contributions with mass $m = 4.389~\gevcc$ and width $\Gamma = 40~\mev$ ($\Gamma = 296~\mev$) in the $\ppbar\eta$ ($\ppbar\omega$) channel with values of $4.03~\textrm{pb}$ and $7.88~\textrm{pb}$ at the $90\%$ CL, respectively. The upper limits for a resonant contribution of the $\psi(4230)$, using current world average values for mass ($m=4.2187~\gevcc$) and width ($\Gamma=44~\mev$)  \cite{pdg} are $7.5~\textrm{pb}$ and $10.4~\textrm{pb}$ at the $90\%$ CL, respectively. 

\section{Summary}

The processes $\ee\to\ppbar\eta$ and $\ee\to\ppbar\omega$ have been studied using $14.7~\textrm{fb}^{-1}$ of electron-positron annihilation data at 17 different center-of-mass energies between 3.7730 GeV and 4.5995 GeV. Both processes are clearly identified at all center-of-mass energies and Born cross sections are determined.
We find no evidence for a  resonant contribution from a fit to the $\ee\to\ppbar\eta$ and $\ee\to\ppbar\omega$ Born cross sections, and set upper limits at the $90\%$ confidence level for a wide range of resonance parameters $m$ and $\Gamma$. Using the approach outlined in Ref.~\cite{Lundborg:2005am}, these upper limits will serve as valuable input for model calculations of the processes $\ppbar\to V\eta$ and $\ppbar\to V\omega$ for the upcoming PANDA experiment.

\begin{acknowledgments}
The BESIII collaboration thanks the staff of BEPCII and the IHEP computing center for their strong support. This work is supported in part by National Key Basic Research Program of China under Contract No. 2015CB856700; National Natural Science Foundation of China (NSFC) under Contracts Nos. 11625523, 11635010, 11735014, 11822506, 11835012, 11935015, 11935016, 11935018, 11961141012; the Chinese Academy of Sciences (CAS) Large-Scale Scientific Facility Program; Joint Large-Scale Scientific Facility Funds of the NSFC and CAS under Contracts Nos. U1732263, U1832207; CAS Key Research Program of Frontier Sciences under Contracts Nos. QYZDJ-SSW-SLH003, QYZDJ-SSW-SLH040; 100 Talents Program of CAS; INPAC and Shanghai Key Laboratory for Particle Physics and Cosmology; ERC under Contract No. 758462; German Research Foundation DFG under Contracts Nos. Collaborative Research Center CRC 1044, FOR 2359, Research Training Group 2149; Istituto Nazionale di Fisica Nucleare, Italy; Ministry of Development of Turkey under Contract No. DPT2006K-120470; National Science and Technology fund; STFC (United Kingdom); The Knut and Alice Wallenberg Foundation (Sweden) under Contract No. 2016.0157; The Royal Society, UK under Contracts Nos. DH140054, DH160214; The Swedish Research Council; Olle Engkvist Foundation under Contract No 200-0605; U. S. Department of Energy under Contracts Nos. DE-FG02-05ER41374, DE-SC-0012069.
\end{acknowledgments}

%

\end{document}